\documentclass[preprint2]{aastex631}

\usepackage[T1]{fontenc}
\usepackage{graphicx}	
\usepackage{amsmath}	
\usepackage{amssymb}	
\usepackage{braket}
\usepackage{float}
\usepackage{xcolor}
\usepackage[normalem]{ulem}

\newcommand{\xhi}{x_\mathrm{HI}}
\newcommand{\tH}{t_\mathrm{H}}
\newcommand{\nh}{n_\mathrm{H}}
\newcommand{\nhi}{n_\mathrm{HI}}
\newcommand{\nhii}{n_\mathrm{HII}}
\newcommand{\trf}{t_\mathrm{rec,full}}

\shorttitle{Ionization Equilibrium \& Impact on LAF Power Spectrum}
\shortauthors{Ku\v{s}mi\'{c} et al.}
\graphicspath{{./}{figures/}}

\begin{document}

\title{Assuming Ionization Equilibrium and the Impact on the Lyman-$\mathrm{\alpha}$ Forest Power Spectrum during the End of Reionization at $8 \geq z \geq 5$}

\author[0000-0002-0761-1985]{Samir Ku\v{s}mi\'{c}}
\affiliation{New Mexico State University, Las Cruces, NM, USA}

\author[0000-0002-0496-1656]{Kristian Finlator}
\affiliation{New Mexico State University, Las Cruces, NM, USA}
\affiliation{Cosmic Dawn Center (DAWN), Niels Bohr Institute, University of Copenhagen / DTU-Space, Technical University of Denmark}

\author{Laura Keating}
\affiliation{Leibniz-Institut f\"{u}r Astrophysik Potsdam, An der Sternwarte 16, 14482
Potsdam, Germany}

\author{Ezra Huscher}
\affiliation{New Mexico State University, Las Cruces, NM, USA}



\begin{abstract}

We explore how the assumption of ionization equilibrium modulates the modeled intergalactic medium (IGM) at the end of the hydrogen Epoch of Reionization using the cosmological radiation hydrodynamic \textsc{Technicolor Dawn} simulation. In neutral and partially-ionized regions where the metagalactic ultraviolet background (UVB) is weak, the ionization timescale $t_\mathrm{ion}\equiv \Gamma^{-1}$ exceeds the Hubble time. Assuming photoionization equilibrium in such regions artificially boosts the ionization rate, accelerating reionization. By contrast, the recombination time $t_\mathrm{rec} < t_\mathrm{ion}$ in photoionized regions, with the result that assuming photoionization equilibrium artificially increases the neutral hydrogen fraction. Using snapshots between $8 \geq z \geq 5$, we compare the predicted Lyman-$\alpha$ forest flux power spectrum with and without the assumption of ionization equilibrium. Small scales ($k > 0.1$ rad s km$^{-1}$) exhibit reduced power from $7 \leq z \leq 5.5$ in the ionization equilibrium case while larger scales are unaffected. This occurs for the same reasons: ionization equilibrium artificially suppresses the neutral fraction in self-shielded gas and boosts ionizations in voids, suppressing small-scale fluctuations in the ionization field. When the volume-averaged neutral fraction drops below $10^{-4}$, the signature of non-equilibrium ionizations on the Lyman-$\alpha$ forest (LAF) disappears. Comparing with recent observations indicates that these non-equilibrium effects are not yet observable in the LAF flux power spectrum.

\end{abstract}



\section{Introduction} \label{sec:intro}

Over the past two decades, a consensus view has emerged in which cosmological hydrogen reionization was driven largely by ultraviolet flux originating in early galaxies~\citep{Bouwens2015,stark2016_galaxies-first-Gyr,Finkelstein2019}, completing around or shortly after $z=6$~\citep{fan06_constraining_ion,becker15_patchy_reion,Planck2020}. While these essentials are now seldom challenged, efforts to use direct measurements of the intergalactic medium's (IGM) physical conditions to test the scenario's predictions remain incomplete because they require accurate treatment of the ways in which the metagalactic ultraviolet background (UVB) propagates through the IGM, ionizes it, and pressurizes it.  This continues to be a major computational challenge despite two decades of effort owing to the prohibitively large range of relevant physical scales  \citep{mcquinn16_igm-evolve}. For example, recent studies have shown that small-scale processes such as the efficiency of star formation in low-mass dark matter halos~\citep{thoul96_gal-formation-sims,Nakatani2020} and photoionization heating of the IGM by passing ionization fronts~\citep{Daloisio2019} remain computational frontiers.

One convenient assumption that is sometimes invoked to simplify IGM models is the idea that the IGM follows photoionization equilibrium~\citep{lidz07_qso-prox_patchy-ion, onorbe17_reion_model_sim, wise19_intro_cosmic_reion, qin2021_reion_gal-inference_LAF}. This approximation is only accurate when both the photoionization and recombination timescales are short compared to a Hubble time and collisional ionizations are negligible. In the diffuse IGM phase that is probed by the Lyman-$\alpha$ forest (LAF;~\citealt{fan06_constraining_ion,mcquinn16_igm-evolve}), the ionization timescale $t_\mathrm{ion}\equiv \Gamma^{-1}$ is very long prior to the epoch of overlap whereas the recombination time, defined as $t_\mathrm{rec,full} \equiv (\alpha_\mathrm{HII} n_H)^{-1}$, grows monotonically owing to cosmological expansion. These considerations indicate that approximations regarding the IGM's ionization state are most likely to yield observable consequences in the LAF. \citet{2019MNRAS.490.1588G} reported disparities in the neutral fractions predicted with versus without ionization equilibrium that were similar in the cases of HI, HeI, and HeII. By contrast, gas near galaxies is more likely to be in ionization equilibrium owing to high density and a locally-enhanced UVB except in rare circumstances where recombinations are inefficient~\citep{Oppenheimer2018}. Our goal is to evaluate directly the extent to which non-equilibrium effects are observable in the IGM by modeling it under two sets of circumstances.

As our baseline case, we adopt the more-accurate density-ionization relationship and the LAF predicted directly by the {\sc Technicolor Dawn} simulation~\citep[TD;][]{finlator18_technicolor-dawn,finlator20_CIV-z>5}, with no approximations added in post-processing other than those that simulate observational effects; we refer to this as the out-of-the-box (OOTB) case. We then re-calculate the hydrogen ionization state of all gas throughout the simulation volume under the assumption of photoionization equilibrium without changing the UVB, gas density, or temperature, referring to this as the ionization-equilibrium (IE) case. By comparing the predictions of the OOTB and IE cases in both physical and observed spaces, we determine how approximations regarding the IGM's ionization state impact a model's accuracy.

In Section \ref{sec:sim_and_analysis}, we review the simulation and tools we use for the analysis. In Section \ref{sec:igm_state}, we explore how the IGM neutral hydrogen fraction evolves during the interval $z=8\rightarrow5$. In Section \ref{sec:laf_ps}, we discuss the use of power spectra analysis and results. In Section \ref{sec:discuss}, we discuss the results and their implications. Throughout this study, the assumed cosmology is a flat $\mathrm{\Lambda}$CDM model with $H_0 = 67.74\: \mathrm{km}\;\mathrm{s}^{-1}\mathrm{Mpc}^{-1},\:X_{\mathrm{H}}= 0.751,\:\Omega_{\mathrm{M}} = 0.3089$, $\Omega_{\mathrm{b}} = 0.0486$, and $\Omega_{\mathrm{\Lambda}} = 0.6911$. We refer to the age of the (modeled) universe as the ``Hubble time" $\tH$, as calculated from the Friedmann equation at the redshift of interest.

\section{Simulations and Analysis} \label{sec:sim_and_analysis}
\subsection{Simulations}\label{ssec:sims}
We anchor our discussion in an analysis of the cosmological radiation hydrodynamic TD simulation \citep[][]{finlator18_technicolor-dawn}. TD is built on \textsc{Gadget-3}, which was last described in~\citet{springel2005}. Gravity is modeled using the default tree-particle-mesh algorithm. Hydrodynamical interactions are modeled using a density-independent formulation of smoothed-particle hydrodynamics~\citep{hopkins2013}. Gas cools radiatively, with primordial cooling computed following~\citet{katz96_reion-sim-treeSPH} and metal-line cooling computed under the assumption of collisional ionization equilibrium~\citep{sutherland1993}. Dense gas acquires a subgrid multiphase interstellar medium~\citep{springel2003} and is converted into collisionless star particles via a Monte Carlo algorithm at a rate that is tuned to reproduce local observations~\citep{springel2003}. Metal enrichment from Type Ia and Type II SNe as well as evolved stars is tracked in 10 independent metal species. Galactic outflows are modeled using a Monte Carlo prescription in which star-forming gas particles receive ``kicks" in momentum space with velocities and rates that are informed by high-resolution simulations~\citep{Muratov2015} and by observations of the galaxy stellar mass function at high redshifts~\citep{finlator20_CIV-z>5}. Our simulation subtends a length of $15h^{-1}\:\mathrm{Mpc}$ ($k=0.419\mathrm{\:rad\:Mpc^{-1}}$) and is the same as in \cite{finlator20_CIV-z>5}; we refer the reader to that work for further details and tests.

Star-forming gas particles emit ionizing flux with a metallicity-dependent efficiency computed from \textsc{Ydggdrasil}~\citep{zackrisson2011} and multiplied by an energy-independent ionizing escape fraction. The propagation of ionizing flux is tracked on-the-fly in 24 independent frequency bins across a regular grid of $80^3$ voxels using an explicit moment method. The discretization's relatively coarse spatial resolution removes the need for approximations regarding the speed of light. The radiation field is attenuated in dense gas (including star-forming regions) under the assumptions of hydrostatic and photoionization equilibrium~\citep{Schaye2001}. The contribution of such self-shielded gas to the metagalactic opacity is likewise reduced. The ionizing escape fraction's dependence on redshift is tuned such that the emergent reionization history simultaneously reproduces observations of the optical depth to Thomson scattering and the mean Lyman-$\alpha$ flux decrement at $z>5$. 

As we focus on a comparison between the results of equilibrium and non-equilibrium ionization, we describe how the IGM's non-equilibrium ionization state is evolved in more detail. At each timestep, the ionization states of hydrogen and helium are advanced using a custom, fully-implicit non-equilibrium solver. Ionization updates account for collisional and photoionization as well as radiative recombinations, but with several exceptions: gas whose temperature exceeds $10^9$K is assumed to be fully-ionized; star-forming gas is assumed to be in ionization equilibrium; and all relevant rates for gas whose temperature falls below 100 K are computed at a temperature floor of 100 K. Very rarely, the solver fails to return a physical solution; when this happens, the relevant particle is assumed to be in ionization equilibrium for that timestep. Updates to the ionization and radiation fields are iterated until the latter has converged. 

Each particle's chemical evolution timescale is computed as the ratio of the electron density to its rate of change $\mathrm{d}t_\mathrm{chem}\equiv 0.5 n_e/\dot{n}_e$. When possible, this timescale is used to regulate the particle's timestep. However, as $\mathrm{d}t_\mathrm{chem}$ can grow prohibitively short within ionization fronts, it is not permitted to drop below 0.25 multiplied by the proper light-crossing time for a single voxel in the radiation field grid. This compromise effectively matches the ionization solver's time resolution with the radiation transport solver's spatial resolution.

\subsection{Analysis}\label{ssec:analysis}
The synthetic Lyman-$\alpha$ forest is extracted along a sightline that traverses the simulation volume at an angle that is oblique with respect to its boundaries. The sightline wraps periodically at simulation boundaries until it subtends a velocity width of $10^6$ km s$^{-1}$, corresponding to an absorption path length of (87.8, 161.8) at $z=(5,8)$ for our cosmology. Our technique for extracting synthetic spectra follows~\citet{Theuns1998} in all respects except that we model absorption as a superposition of Voigt profiles \citep{Humlicek1979} rather than Gaussians. Synthetic pixels span a velocity width of 2.5 km s$^{-1}$ pixel$^{-1}$. The spectrum is smoothed with a Gaussian with a full width at half-maximum of 6 km s$^{-1}$ to model instrumental broadening. Finally, Gaussian noise is added to each pixel assuming signal-to-noise of 20 per pixel.

At each redshift, we extract a simulated spectrum twice. First, we use the IGM ionization field directly predicted by the simulation, accurately capturing the impact of non-equilibrium ionizations on the Lyman-$\alpha$ forest. For comparison, we then re-compute each particle's ionization state under the assumption of photoionization equilibrium and re-extract the sightlines. Gas temperatures, densities, and proper motions are identical in the two cases, and the radiation field is likewise unchanged.

\begin{figure}
    \centering
    \includegraphics[width=0.5\textwidth]{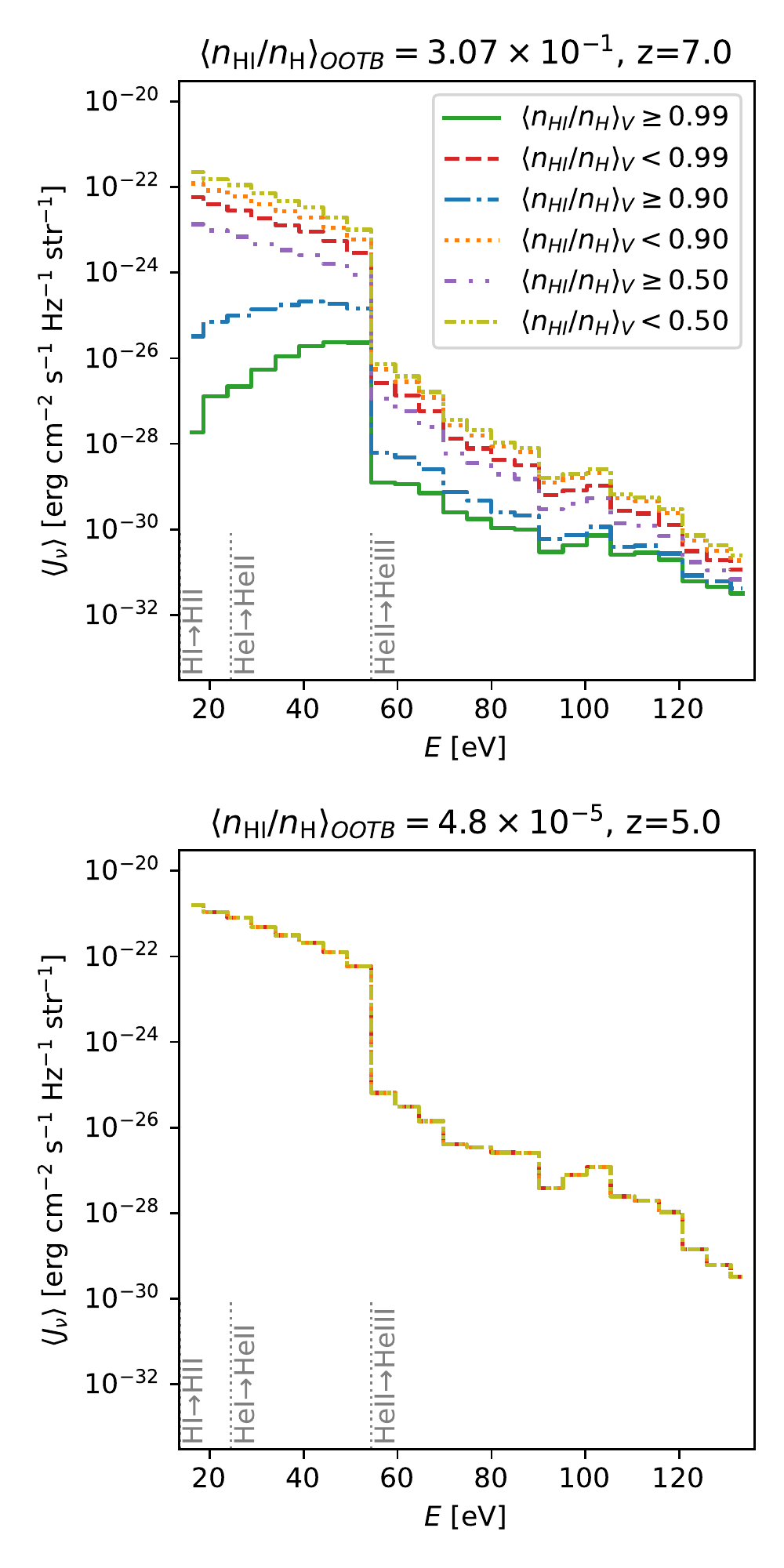}
    \caption{Neutral fraction-weighted SED of the UVB of different regions at the redshifts $z=7$ and $z=5$. The multiple lines are of regions that fit the specific neutral fraction criteria presented in the legend, so one region can be represented by multiple lines. Ionization lines for various species expected in the IGM annotated in bottom plot. In the bottom plot, all lines but $\langle \nhi / \nh \rangle_V < 0.5$ is plotted, showing that all regions, by volume average, are majority ionized by $z=5$. This plot shows considerable absorption of the UVB by hydrogen in more $\mathrm{H}$ neutral regions, alongside the absorption for the various ionizations of $\mathrm{He}$. }
    \label{fig:nu_Jnu}
\end{figure}

\begin{figure*}
    \centering
    \includegraphics[width=0.49\textwidth]{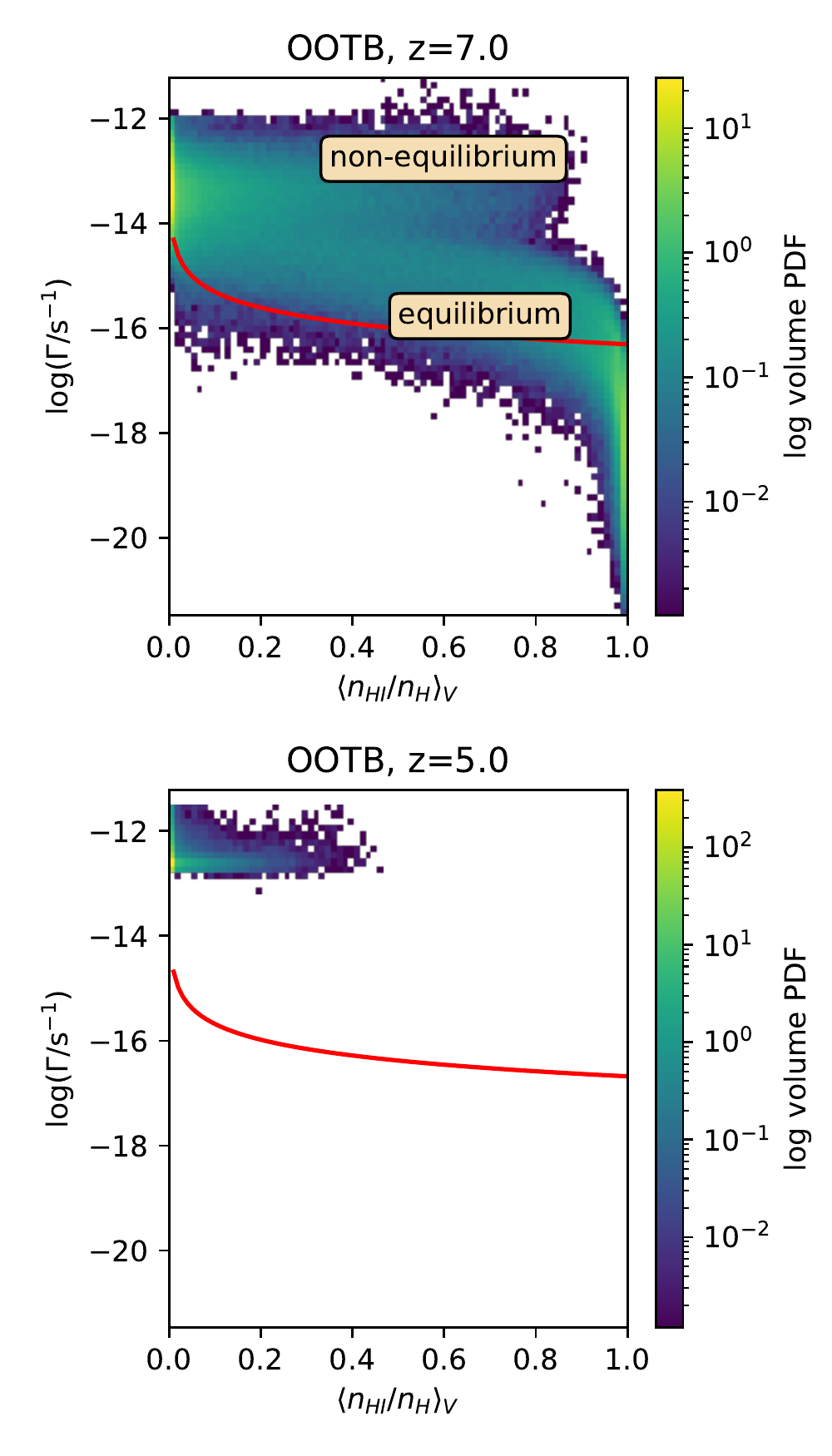}
    \includegraphics[width=0.49\textwidth]{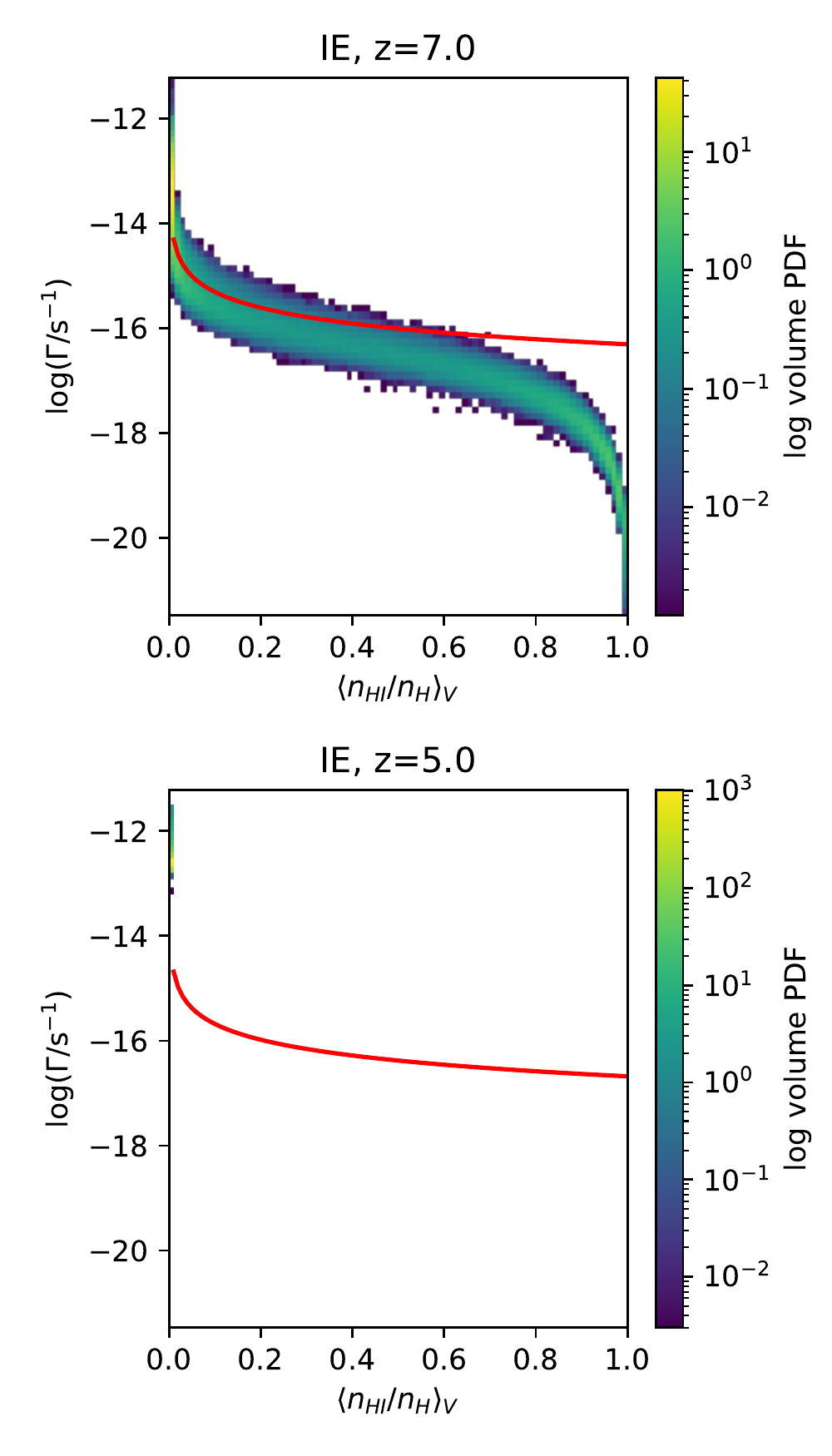}
    \caption{2D histogram showing the distribution of voxel regions by volume-weighted neutral fraction and its ionization rate $\Gamma$ of non-equilirbium (OOTB) and ionization equilibrium (IE) neutral fractions. Non-equilibrium quantities were taken directly from the simulation outputs, while the equilibrium neutral fraction calculated the neutral fraction from simulation outputs using Equation \ref{eqn:ion-eq}}. The red, solid line shows the possible solutions for Equation \ref{eqn:ion-eq} using the critical density at temperature $T=10^4\:\mathrm{K}$. At $z \geq 7$ a bimodality is present in the OOTB for partially ionized regions, where higher $\Gamma$ are regions with ionization fronts and the other are not. This confirms by $z=5$ that all regions have $\langle \nhi / \nh \rangle_V < 0.5$. The lower branch in the OOTB case are gases that are close to photoionization equilibriumas seen in the trend.
    \label{fig:xH_Gamma}
\end{figure*}

\begin{figure*}
    \centering
    \includegraphics[width=\textwidth]{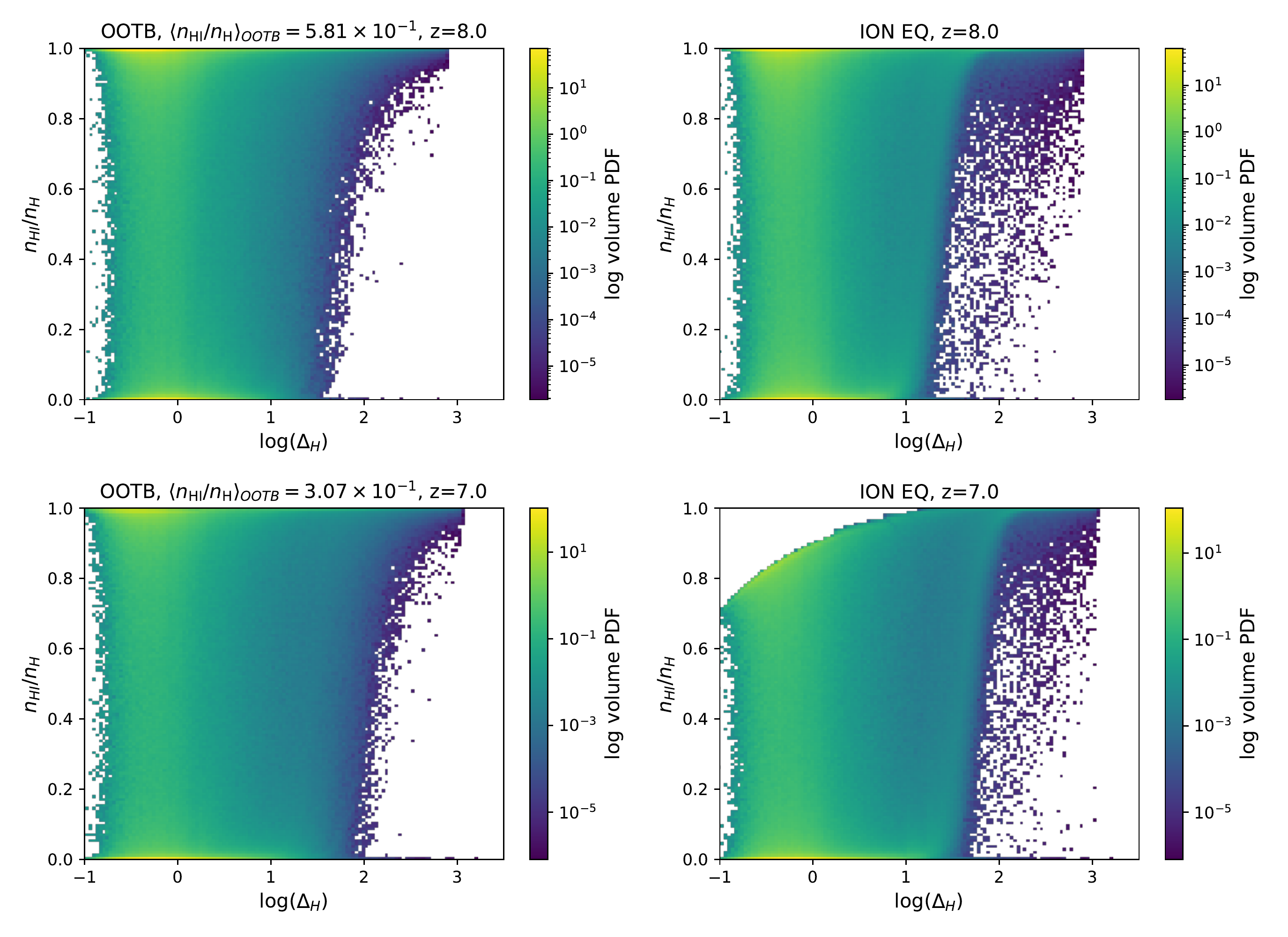}
    \caption{\label{fig:HI_8-7} Neutral fraction of hydrogen plotted against the overdensity of the hydrogen in our simulation comparing the ionization equilibrium case against the OOTB case. Plotted here are from snapshots taken at $z=8$ and $7$. White-space denotes no data points. During these times, hydrogen gas under ionization equilibrium appears more concentrated with ionized gas than in our OOTB case. Note the apparent forbidden region in ionization equilibrium hydrogen, where low-density gas cannot be of a certain neutral fraction.}
\end{figure*}

\begin{figure*}
    \centering
    \includegraphics[width=\textwidth]{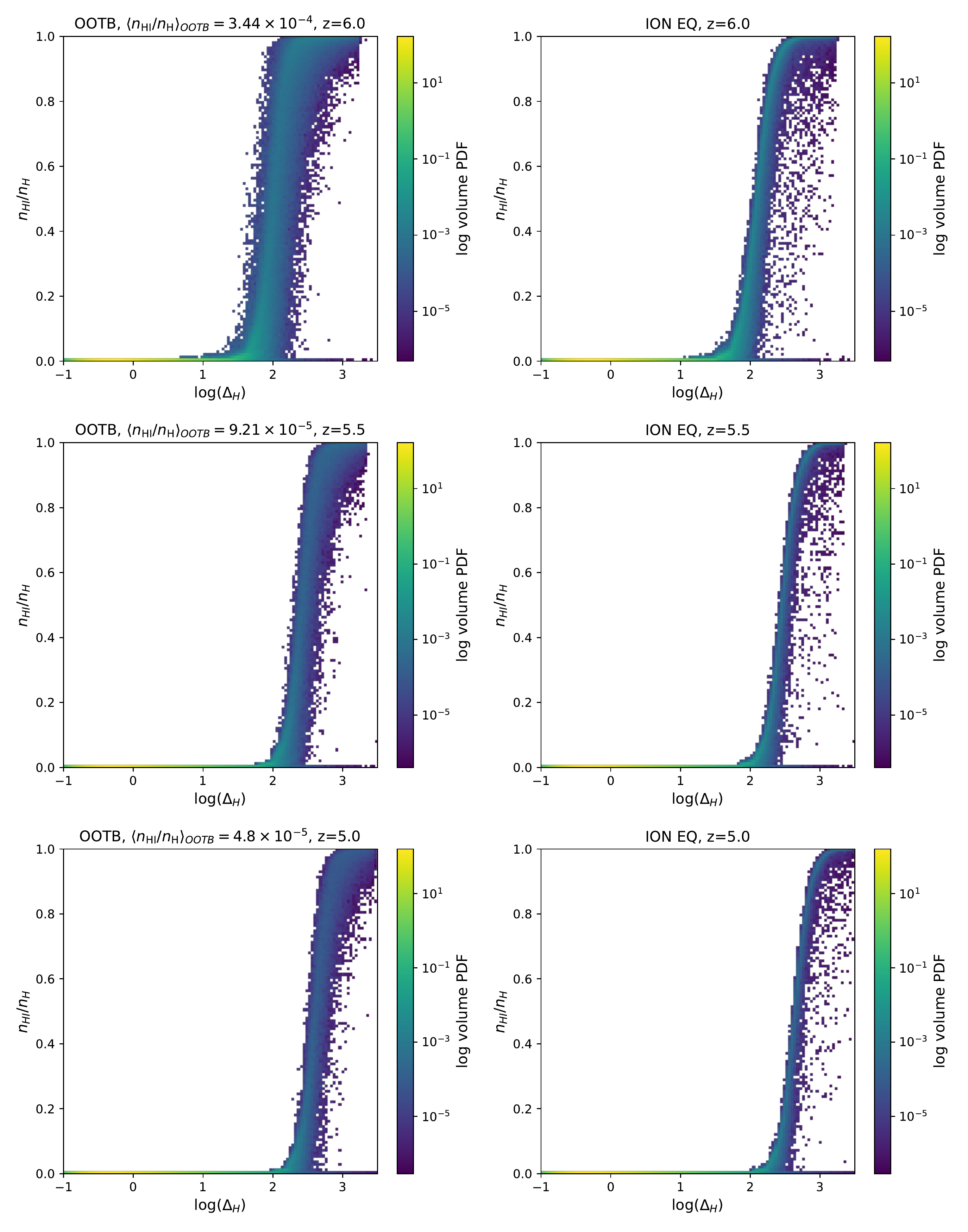}
    \caption{\label{fig:HI_6-5} Similar plot as Figure \ref{fig:HI_8-7}, but now looking at snapshots at $z=6$, $5.5$, and $5$. Ionization equilibrium not only appears to have higher concentration of ionized hydrogen gas per parcel, but also more closely follows a potential fit for an ionization fraction model than our OOTB case. There is a larger scatter in neutral fraction for a given overdensity in the OOTB case.}
\end{figure*}

\section{IGM Thermal State} \label{sec:igm_state}
Well after the completion of reionization, the highly-ionized IGM obeys a simple relationship between photoionization rate per hydrogen atom $\Gamma$, neutral hydrogen fraction $x_\mathrm{HI}\equiv n_\mathrm{HI}/n_\mathrm{H}$, gas density $n_H$, and recombination coefficient $\alpha$~\citep{mcquinn16_igm-evolve}:
\begin{equation} \label{eqn:ion-eq}
x_\mathrm{HI} = \frac{\alpha n_H}{\Gamma}.
\end{equation}
We compute $\Gamma$ as
\begin{equation}
    \Gamma \equiv \int_{\nu_\mathrm{LL}}^{\infty} \frac{4 \pi J(\nu)}{h \nu} \sigma_{\mathrm{LL}}(\nu) d\nu,
\end{equation}
where $\nu_{\mathrm{LL}}$ and $\sigma_{\mathrm{LL}}(\nu)$ are the HI ionization threshold frequency and cross-section \citep{katz96_reion-sim-treeSPH}. Departures from the assumptions of photoionization equilibrium and a homogeneous UVB (as quantified by $\Gamma$) therefore manifest as large-but-shrinking scatter in the relationships between $x_\mathrm{HI}$, $\Gamma$, and $n_H$. In this section, we verify that the predicted scatter is significant and quantify departures from ionization equilibrium.

\subsection{UV Background} \label{sub:uvb}
The primary driver of the IGM's ionization state is the UVB \citep{mcquinn16_igm-evolve}. As such, understanding the UVB's spatial inhomogeneity and its relationship to the gas density field is a first step to understanding the IGM's evolution. The UVB is left unchanged when we re-compute the IGM neutral fraction under the assumption of ionization equilibrium, hence we will show that it drives small-scale opacity fluctuations that are only captured accurately when the assumption of ionization equilibrium is relaxed. Our simulation's self-consistent, spatially-inhomogeneous UVB provides improved realism with respect to results from assuming a homogeneous UVB.

In the top panel of Figure \ref{fig:nu_Jnu}, we show that the UVB's slope and normalization both vary dramatically with the local neutral hydrogen fraction at $z=7$, when the volume-averaged neutral fraction is $\approx31\%$. In particular, an increase in the local neutral fraction significantly increases absorption near the hydrogen ionization edge while permitting some high-ionization flux to penetrate. This spectral filtering is a well-known characteristic of ionization fronts \citep{Abel1999,Daloisio2019}. 
The fact that even voxels that remain highly neutral do contain a small amount of ionizing flux (see, for example, the solid green curve) may, to some extent, be an artifact of our moment-based radiation transport solver because moment methods are fairly diffusive. On the other hand, it is not unrealistic given that regions completely devoid of high-energy flux should be rare by $z=7$. For example, the mean distance between galaxies with absolute magnitude brighter than $M_\mathrm{UV}=-15$ is observed to be $\approx2$ comoving Mpc \citep[adopting an extrapolation of][]{Bouwens2021}. This distance roughly matches the mean free path in the neutral IGM for light with energies at twice the HI ionization edge; it falls below the mean free path at higher energies. Hence it is reasonable to expect that many if not most neutral regions will be permeated by a weak but spectrally-hard UVB by $z=7$.

The inhomogeneous UVB, when combined with a non-equilibrium ionization history, gives rise to significant scatter in the relationship between the hydrogen photoionization rate $\Gamma$ and the local neutral fraction at $z=7$. We show this relationship at $z=7$ and $z=5$ in Figure \ref{fig:xH_Gamma} for the OOTB (left) and IE (right) cases. Both quantities are averaged over the $187.5 h^{-1}$ comoving kpc voxels, which suppresses small-scale UVB fluctuations. The top-right (IE) panel confirms that, prior to overlap, $\Gamma$ declines with increasing neutral fraction. A similar trend is visible in the OOTB case's lower branch at $z=7$. Moreover, although the scatter in $\Gamma$ at fixed neutral fraction is large in the OOTB case, the lower end of the scatter agrees with the IE case; this boundary indicates that a minimum neutral fraction exists for any value of $\Gamma$, corresponding roughly to mean-density gas. The relationship's large scatter reflects both the inhomogenous UVB and the non-equilibrium ionization, leading to higher possible neutral fractions in the OOTB case where the IE case would accelerate reionization. 

In the OOTB case, there is a clear bimodality in $\Gamma$ such that regions at constant neutral fraction can have either a high $\Gamma$ or a lower $\Gamma$. This likely arises because the higher branch traces regions around ionization fronts. Since the lower branch more closely follows ionization equilibrium (red curve), this suggests that the higher, more-neutral branch is gas that is not in ionization equilibrium because the ionization timescale is long. Comparing Figure \ref{fig:xH_Gamma} with Figure \ref{fig:nu_Jnu}, we find that the UVB amplitude is elevated throughout this non-equilibrium branch irrespective of neutral fraction. This necessitates a qualification to Figure~\ref{fig:nu_Jnu}: the UVB is only suppressed in neutral regions that are in ionization equilibrium, not in ionization fronts. The bottom panel shows that, by $z=5$, $\Gamma$ becomes independent of the neutral fraction and no regions are more than half-neutral. This is consistent with its higher UVB intensity, fitting criteria that show it is less than half-ionized, and lacking the HI ionization absorption feature in Figure \ref{fig:nu_Jnu}; it is what is expected of an ionized universe after reionization. However, due to the averaging within the voxels, there are fully neutral gas particles, but are filtered when looking at these scales comparable to halos. At both redshifts, the predicted scatter is larger in the OOTB case than in the IE case.

\subsection{Overdensity} \label{sub:overdense}
The bimodal distribution of $\Gamma$ as a function of neutral fraction in Figure~\ref{fig:xH_Gamma} signifies a departure from full ionization equilibrium, which can also drive scatter in the relationship between density and neutral fraction. Put differently, the assumption of ionization equilibrium artificially suppresses scatter in the relationship between ionization fraction and gas density. In order to quantify this effect, we show in
Figures \ref{fig:HI_8-7} and \ref{fig:HI_6-5} the evolving relationship of the neutral fraction to the hydrogen overdensity $\Delta_{\mathrm{H}} \equiv \rho/\langle \rho \rangle$ in the OOTB and IE cases. At $z=8$, regions with hydrogen overdensities, i.e., within 1 dex of the mean density, show no clear relationship between ionization state and density, indicating that the UVB remains far too weak for its environmental dependence (Figure~\ref{fig:nu_Jnu}) to create observational signatures, at least on the scales that we consider. Moving to higher overdensities where gas traces filaments and halos, the regions grow increasingly neutral owing to self-shielding. At $z=7$ we see two interesting effects: (1) ionization fronts have increasingly penetrated filamentary regions ($\Delta_{\mathrm{H}} > 1$), reflecting the final stage of reionization \citep{Finlator2009}; and (2) there is a conspicuous maximum neutral fraction in and around voids in the IE case that is missing in the OOTB calculation: starting at $\Delta_{\mathrm{H}} < 1$, the IGM's neutral fraction displays a ``ceiling" that grows with density. Once again, IE artificially boosts ionizations in diffuse, optically-thick gas where $\Gamma$ is small but nonzero. Afterwards at $z\leq6$ the cases converge, although the IE case still yields a slightly tighter relationship between overdensity and neutral fraction and artificially suppresses the neutral fraction in collapsed regions ($\Delta_{\mathrm{H}}>200$), even at $z=5$. In short, imposing IE artificially increases the ionization fraction wherever the UVB is weak. 

Figures~\ref{fig:HI_8-7}--\ref{fig:HI_6-5} suggest that IE accelerates reionization. Although we do not demonstrate this directly by re-running our simulation in the IE approximation~\citep[cf.][]{2019MNRAS.490.1588G}, we confirm this qualitatively by recomputing the volume-averaged neutral fraction at each redshift and between the two cases. Taking each $i$th particle and using the specific volume $V_i \equiv 1/\rho_i$, we compute the volume-weighted mean neutral fraction as

\begin{equation}
    \langle n_{HI}/n_H \rangle_{V} = \frac{\sum_i^N V_i (n_{HI}/n_H)_i}{\sum_i^N V_i}
\end{equation}

\begin{figure}
    \centering
    \includegraphics[width=0.5\textwidth]{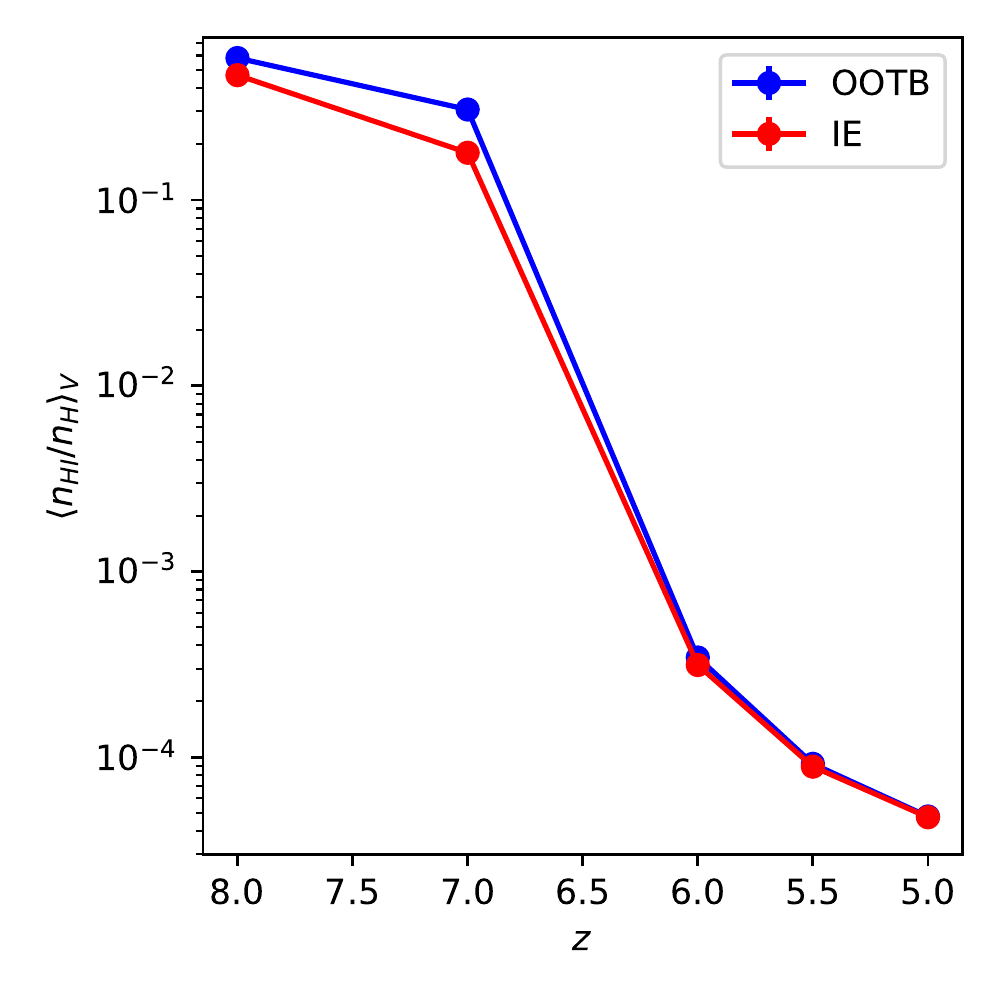}
    \caption{\label{fig:mean_fHI} Volume-weighted mean neutral fraction of hydrogen at each redshift, comparing the differences between the OOTB (blue) and IE (red) cases. Throughout $8 \leq z < 5$, hydrogen in the IE case is more ionized than in the OOTB case. The neutral fractions converge by $z=5$.}
\end{figure}

\begin{table}
\centering
\caption{Ratio of the volume-weighted neutral fractions in each redshift $\langle n_{\mathrm{HI}} / n_{\mathrm{H}} \rangle_{OOTB} / \langle n_{\mathrm{HI}} / n_{\mathrm{H}} \rangle_{IE}$. This focuses on particles that are either in the void ($\Delta < 1$) or halo ($\Delta > 200$) regions.}
\label{tab:fHI_ratios}
\begin{tabular}{rrr}
\hline
 $z$ &  Voids &  Halos \\
\hline
 8.0 &   1.22 &   1.01 \\
 7.0 &   1.73 &   1.00 \\
 6.0 &   1.02 &   1.00 \\
 5.5 &   1.01 &   1.06 \\
 5.0 &   0.99 &   1.05 \\
\hline
\end{tabular}
\end{table}

In Figure \ref{fig:mean_fHI} we plot the volume-weighted neutral-fraction for each redshift between the two cases. Here, we can see the general trends hidden within Figures \ref{fig:HI_8-7} and \ref{fig:HI_6-5}. The gas under IE is 19\% and 41\% more ionized than in our OOTB case at $z=8$ and $z=7$ respectively. Following the completion of reionization, the two neutral fractions converge to values in the range $\langle n_{HI}/n_H \rangle_{V} \sim 5 \times 10^{-5}$. The impact of imposing IE on the topology of reionization emerges from comparing the neutral fractions in voids and in halos ($\Delta_{\mathrm{H}}<1$ and $\Delta_{\mathrm{H}}>200$, respectively). We show the ratio of these quantities in Table \ref{tab:fHI_ratios}. A ratio that is ($>1$, $<1$) indicates that reionization is (more, less) complete in the IE case. IE evidently accelerates reionization in voids, with the two cases roughly converging by $z=5$. By contrast, the neutral fraction in halos is less sensitive to the treatment of ionizations. We conclude that the ionization state of dense gas more nearly obeys ionization equilibrium than voids even when this is not imposed despite the fact that our self-shielding treatment qualitatively boosts its ionization timescale. On the other hand, the fact that its ionization timescale is non-trivial still causes the IE case to over-ionize dense gas at $z<6$ by allowing ionizing radiation to penetrate further into a cloud. In the next section, we analyze these timescales in more detail.

\subsection{Timescales} \label{sub:timescales}

In the previous section, we showed that imposing ionization equilibrium on the IGM in post-processing artificially suppresses the neutral fraction with respect to the OOTB case during the EOR either by IE accelerating ionizations in regions where the ionizing background is weak or else by OOTB accelerating recombinations in gas that is hot or diffuse. In order to verify that these effects stem from mismatched timescales, we compute the ionization timescale $t_\mathrm{ion}$ and the recombination timescale $t_\mathrm{rec}$ at different times. For consistency with the simulation, we compute the temperature-dependent HII recombination rate following Table 2 of \cite{katz96_reion-sim-treeSPH}.

\begin{figure*}
    \centering
    \includegraphics[width=\textwidth]{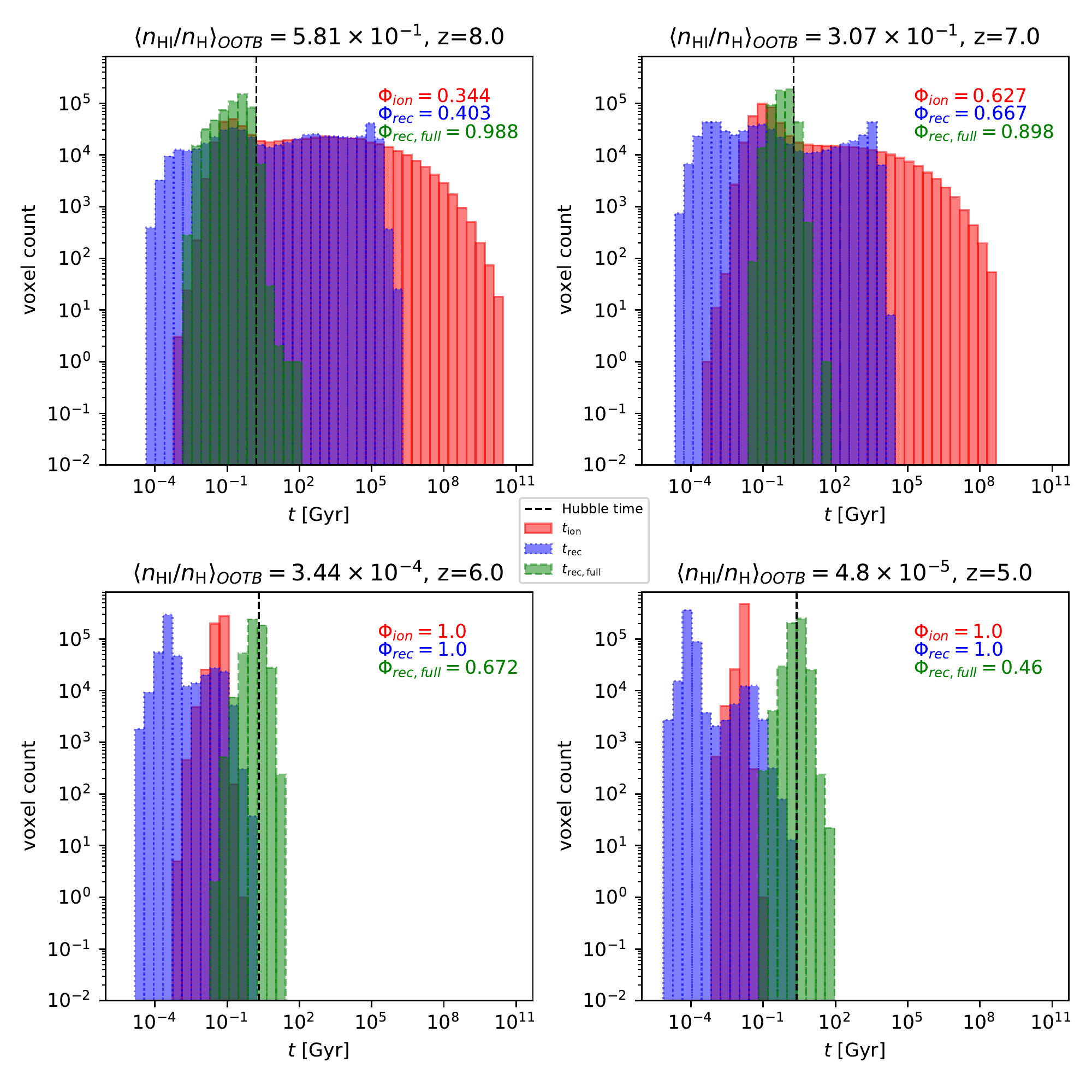}
    \caption{\label{fig:trec_hist} The distributions of recombination and ionization timescales at each redshift when averaging spatially over voxels of constant size 187.5 h$^{-1}$ comoving kpc. Both the ionization timescale (Equation~\ref{eqn:tion}; red, solid) and the recombination timescale (Equation~\ref{eqn:tRecombPart}; blue, dotted) are compared to the condition to recombine from completely ionized to completely neutral (Equation~\ref{eqn:tRecombFull}; green, dashed). The vertical dashed, black line is the Hubble time at its respective redshift. The legends show the fraction $\Phi$ of voxels whose timescales are less than the Hubble time. Prior to overlap ($z>6$), both the ionization and recombination timescales exceed the Hubble time for more than half of the universe. By $z=5$, recombination and ionization timescales both fall below the Hubble time. Meanwhile, the timescale for full recombination continues to grow owing to cosmological expansion, with only $\sim 46\%$ of the universe dense enough to recombine fully within a Hubble time by $z=5$.}
\end{figure*}

To use the information necessary for the recombination timescales, we extracted the ionization rate parameter $\Gamma$, local neutral hydrogen fraction, temperature, and hydrogen density for each voxel in each redshift. In order to derive meaningful definitions for ionization and recombination timescales, we begin with the differential equation for the change in neutral hydrogen number density $\nhi$:

\begin{equation}
    \frac{d}{dt} \nhi = -\Gamma \nhi + \alpha(T) n_e \nhii - k_{e}n_e\nhi
\end{equation}
\noindent
where $\Gamma$ is the photoionization rate, $\alpha(T)$ is the radiative recombination rate coefficient, $n_e$ is the electron number density, and $k_{e}$ is the collisional ionization rate coefficient. Neglecting recombinations and collisional ionizations yields the ionization e-folding time $t_\mathrm{ion}$:
\begin{equation}\label{eqn:tion}
    t_\mathrm{ion} \equiv \frac{1}{\Gamma}
\end{equation}
If $t_\mathrm{ion} > t_\mathrm{H}$, then photoionizations occur slowly and the IGM is more neutral than one infers from the assumption of ionization equilibrium.

In order to derive the timescale for gas to recombine from being fully-ionized to the current ionization state assuming the current recombination rate, we substitute the number densities with respect to the neutral fraction $\xhi$: $\nhi \rightarrow \xhi\nh$ and $\nhii \equiv (1-\xhi)\nh$. With these substitutions, we define $t_\mathrm{rec}$ as

\begin{equation}\label{eqn:tRecombPart}
    t_\mathrm{rec} \equiv \frac{\xhi}{(1-\xhi)^{2} \alpha(T) n_{H}}.
\end{equation}
If $t_\mathrm{rec} > \tH$, then gas cannot recombine from fully-ionized to its \emph{current} ionization state in a Hubble time; such gas, once ionized, will remain so. In order to accommodate the possibility of fully-neutral gas, we may alternatively define the recombination timescale as the time to recombine completely from an initial neutral fraction of 0, which we label $\trf$:

\begin{equation}\label{eqn:tRecombFull}
    \trf \equiv \frac{1}{\alpha(T) n_{\mathrm{H}}}
\end{equation}
Note that, in the highly-ionized case, IE can be expressed as the condition $\xhi=\trf/t_\mathrm{ion}$ (Equation~\ref{eqn:ion-eq}).
For gas that is at least partially-ionized, $t_\mathrm{rec} < \trf$, hence it is possible that ionized gas for which $\trf > \tH$ is still nearly in ionization equilibrium and this calculation provides a ceiling to the recombination timescale. Using these timescales, we now illustrate how the IGM departs from ionization equilibrium at different times. Broadly, any process whose timescale is shorter than $\tH$ can safely be assumed to be near equilibrium.  

In Figure \ref{fig:trec_hist}, we show the distribution of voxel-averaged timescales at four different redshifts. We additionally compute the volume fraction $\Phi$ of the universe for which each process is in equilibrium and show these fractions in the legends. At $z=8$, the majority of the universe is not yet in equilibrium: The ionization timescale (red) vastly exceeds $\tH$ wherever ionization fronts have not yet passed. $t_\mathrm{rec}$ also exceeds $\tH$ in neutral regions because free protons and electrons are scarce (blue). However, regions that are ionized recombine efficiently (green). At $z=7$, the situation is similar although the distributions of timescales are narrower because the longest ionization and recombination timescales are no longer populated. This shows that the evolution rate reduces in each region. By $z=6$, $t_\mathrm{ion}<\tH$ due to the increasing amplitude of the UVB. Likewise, $t_\mathrm{rec} < \tH$ because free electrons and protons grow abundant and $\xhi \ll 1$. These developments indicate that ionization equilibrium eventually applies to the majority of the universe. Trends at $z=5$ are largely indistinguishable from $z=6$.

When averaging over scales that are large compared to collapsed structures, $\trf$ is small and grows slowly owing to cosmological expansion. Once star formation begins, $t_\mathrm{ion}$ falls below $t_H$ for a growing volume fraction, pushing the universe closer to the IE case. Even so, however, $t_\mathrm{ion} > t_\mathrm{rec}$ throughout much of the universe. 

On smaller scales, $\trf>\tH$ in hot, diffuse regions such as virial shocks and regions where gas has been heated by star formation feedback. Such regions generically lie close to star formation sites where the UVB is locally strong, hence IE will qualitatively accelerate their ionizations, making them too ionized. This is indeed seen in Table \ref{tab:fHI_ratios} at $z=8$ to $7$, but not so much at later times.

Both of these considerations explain why the OOTB case is generally more neutral than the IE case over the entire universe, as seen in Figure~\ref{fig:mean_fHI}.

\section{Lyman-$\alpha$ Forest Power Spectrum} \label{sec:laf_ps}

\begin{figure*}
    \centering
    \includegraphics[width=\textwidth]{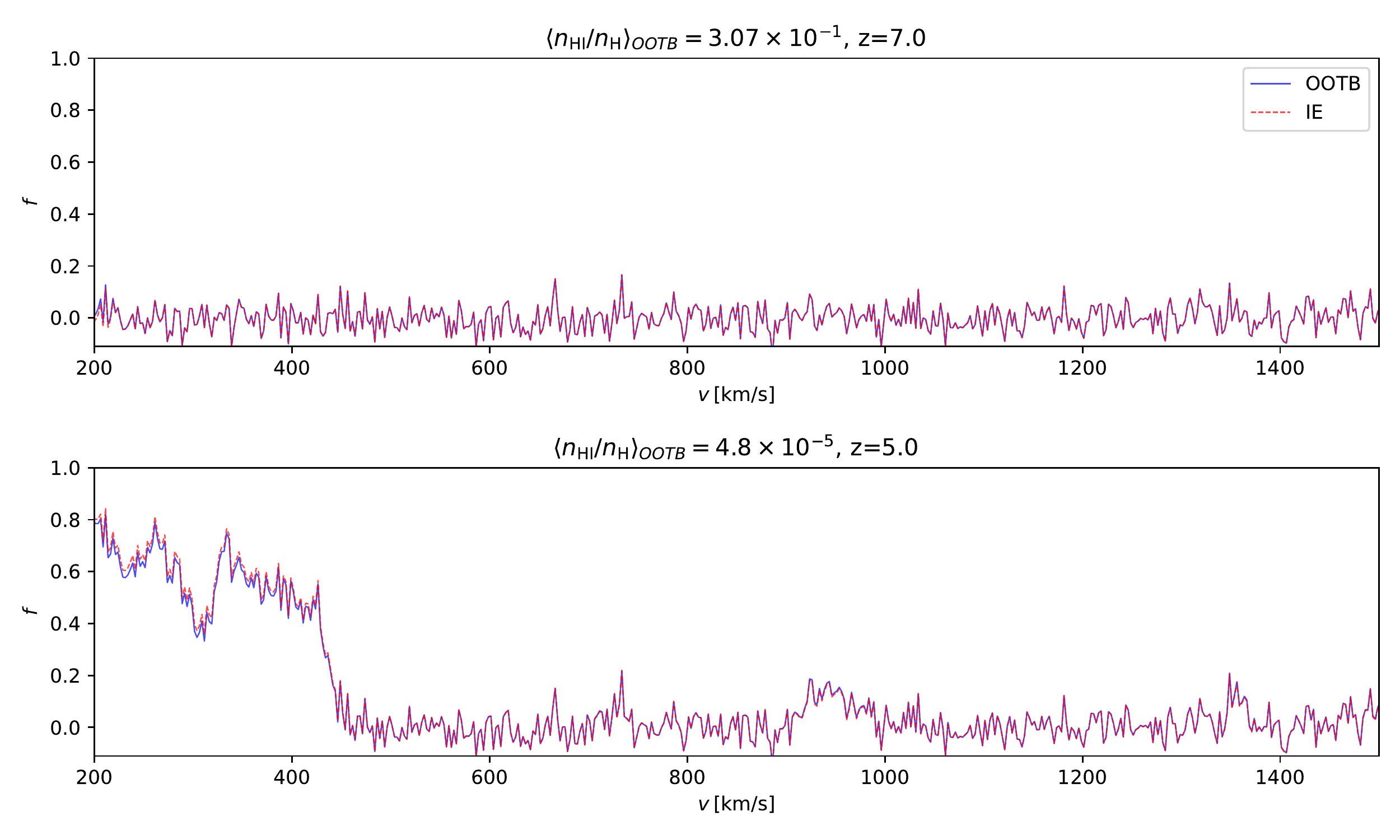}
    \caption{Part of the simulated spectra of the LAF at $z=7$ and $5$ of both the OOTB and IE cases. The flux density is normalized to the continuum. Slight differences are noticeable in the flux values between the two cases, but these differences are not constant throughout.}
    \label{fig:laf_spectra}
\end{figure*}

\begin{figure*}
    \centering
    \includegraphics[width=0.49\textwidth]{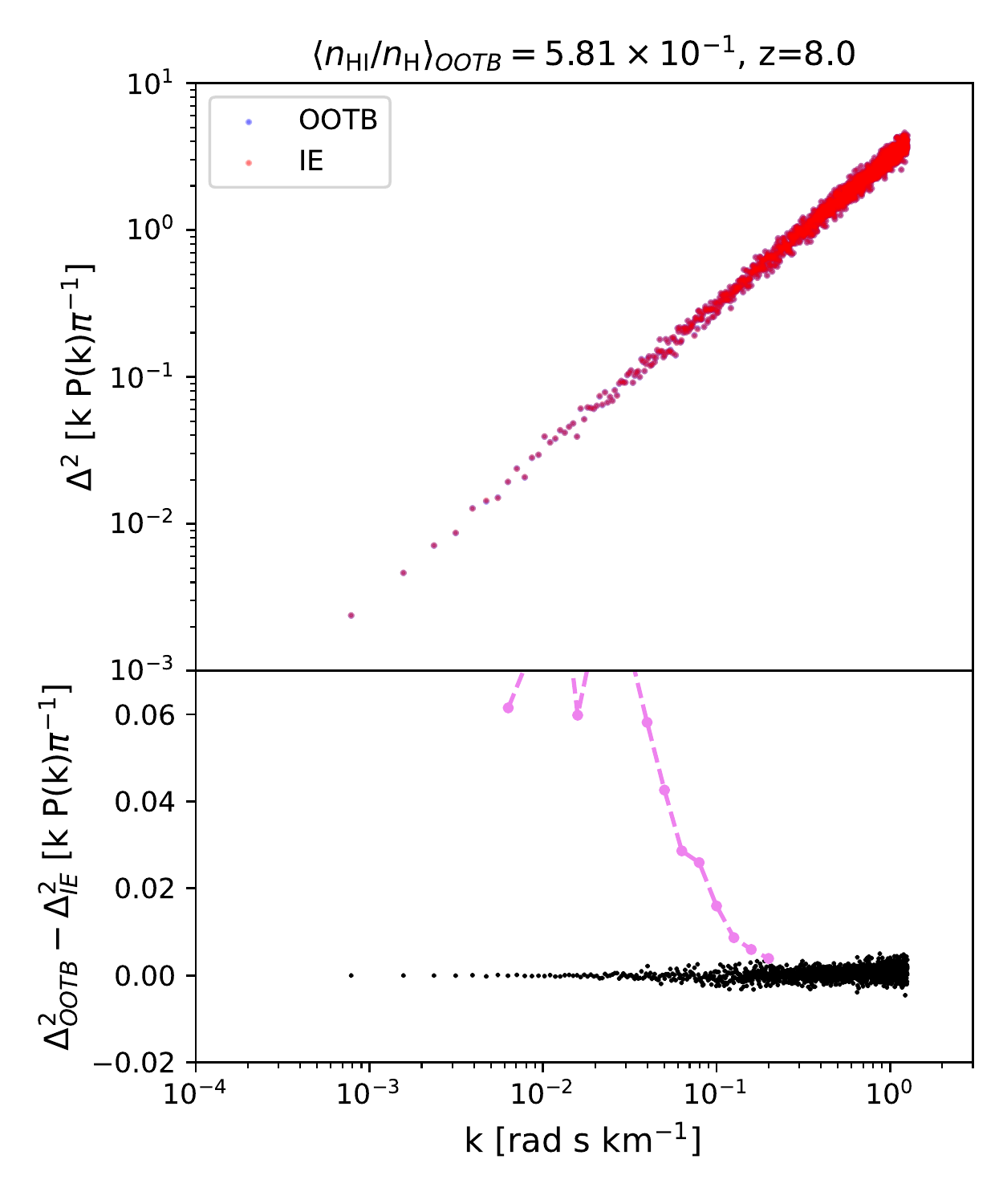}
    \includegraphics[width=0.49\textwidth]{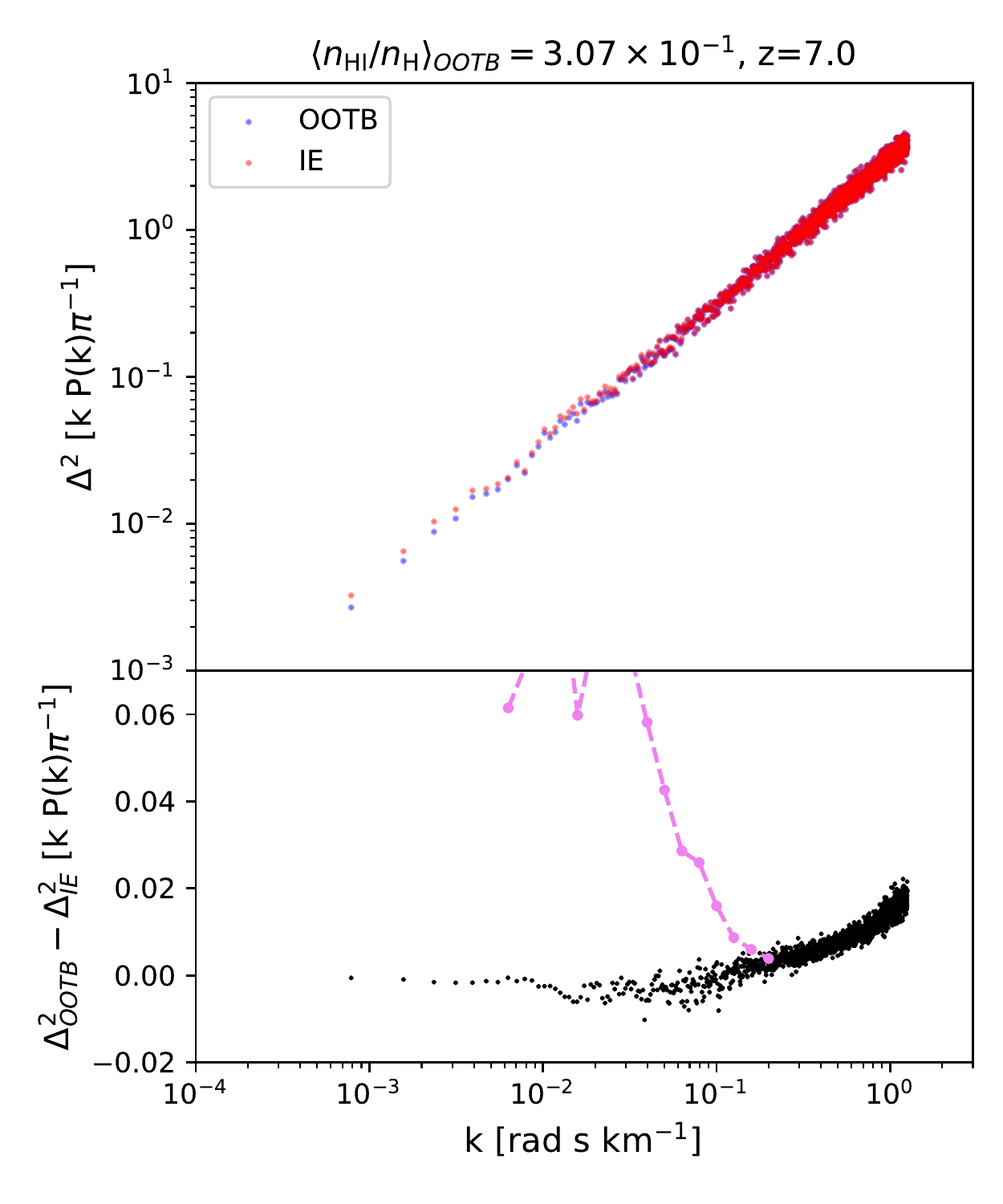}
    \\
    \includegraphics[width=0.49\textwidth]{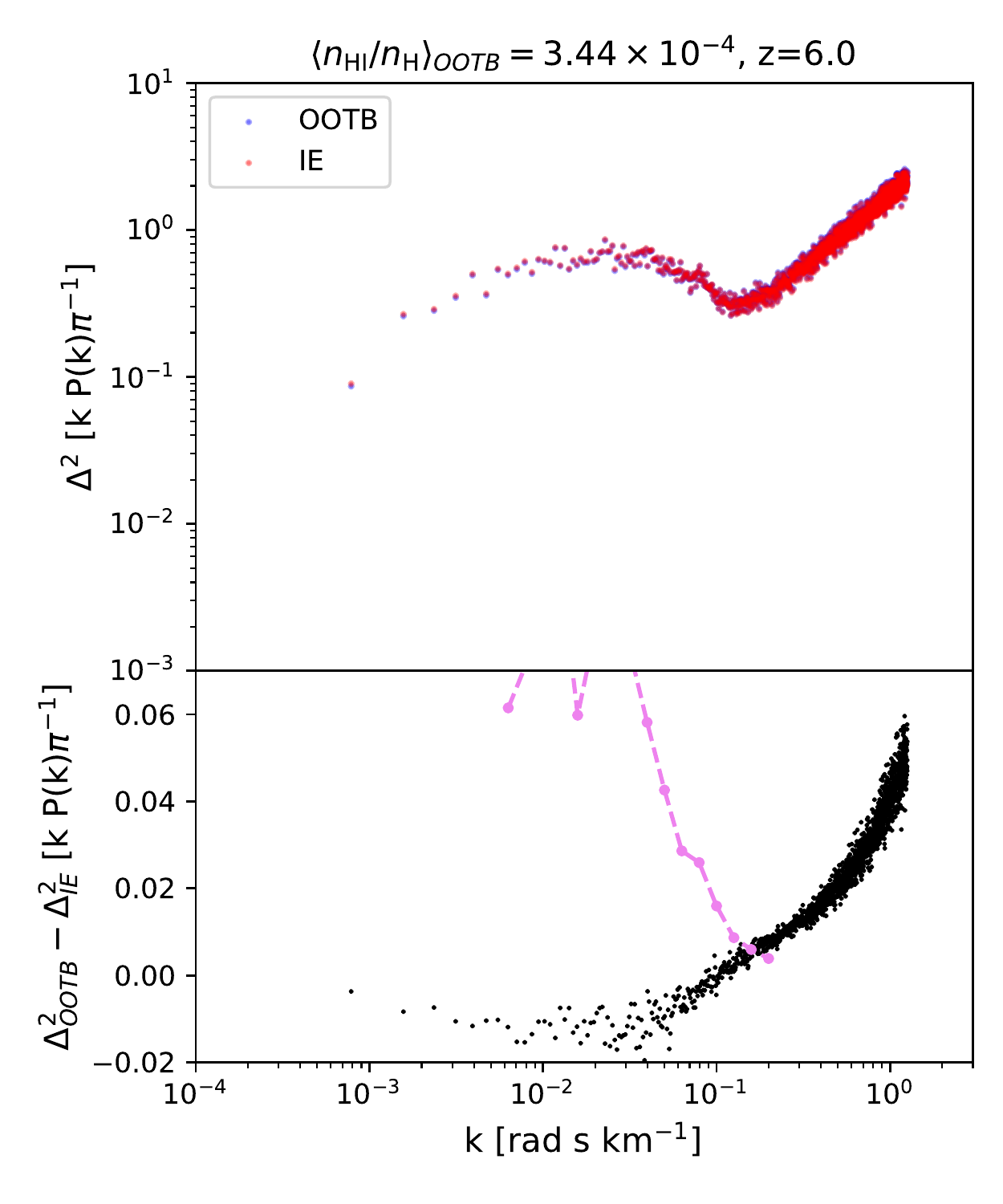}
    \includegraphics[width=0.49\textwidth]{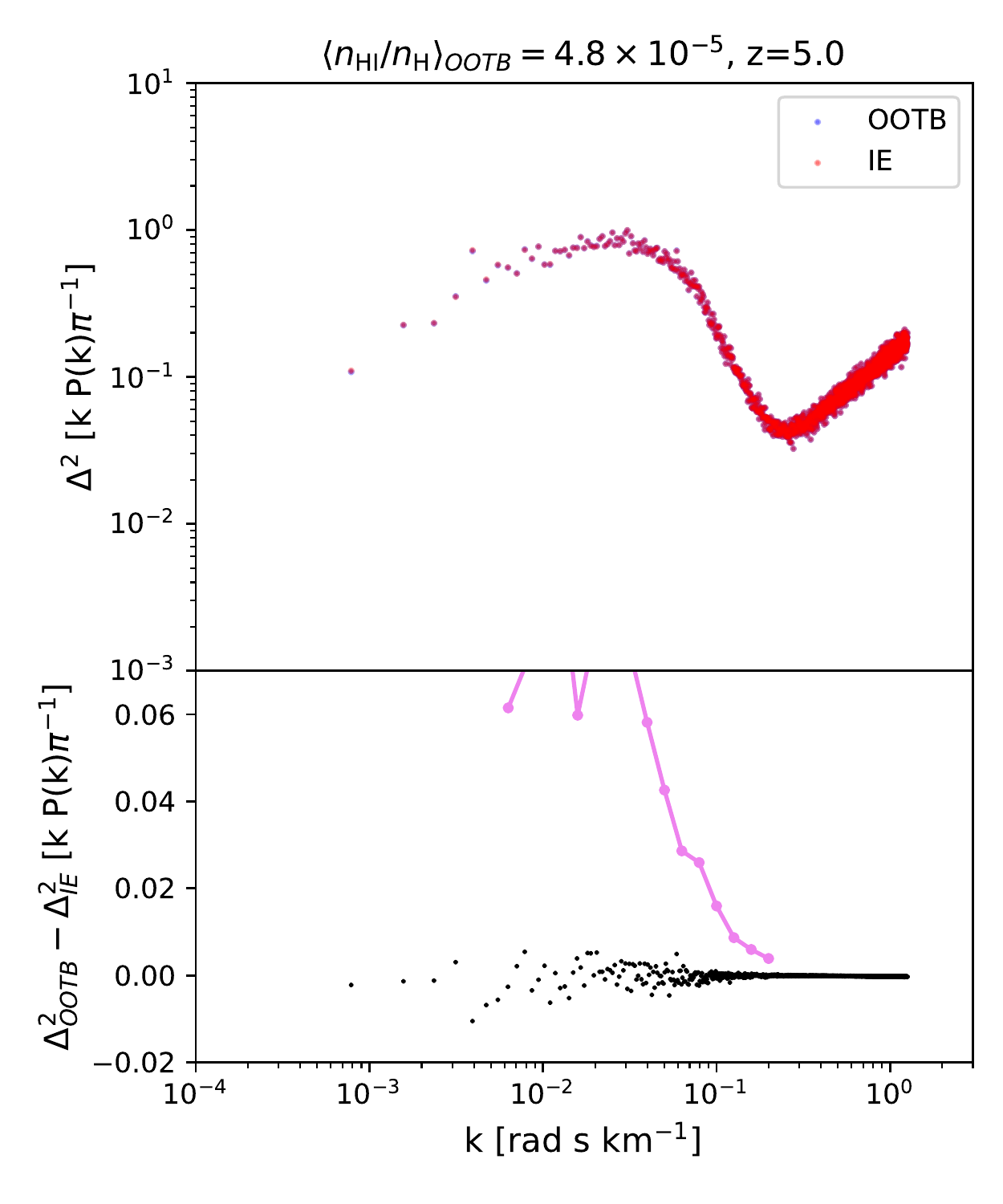}
    \caption{ Power spectra of the LAF for both the OOTB case and the ionization equilibrium case and their differences. Each plot is of a different snapshot in redshift. Both follow a trend of reduced power in smaller-scale structures as redshift decreases. From $7 \leq z < 5$, we see reduced power in small-scale structures for our LAF under the ionization equilibrium case. The violet line are $1\sigma$ uncertainties in the power spectrum from observations done by \cite{boera19_PS-constraints}. Since the uncertainties are from the power spectrum at $z=5$, it is kept solid in the snapshot of the same redshift, but plotted as dashed at higher redshifts in order to compare expected observability. This seems expected, as we see less neutral hydrogen than in the OOTB case.}
    \label{fig:PS_almost_all}
\end{figure*}

\begin{figure}
    \centering
     \includegraphics[width=0.5\textwidth]{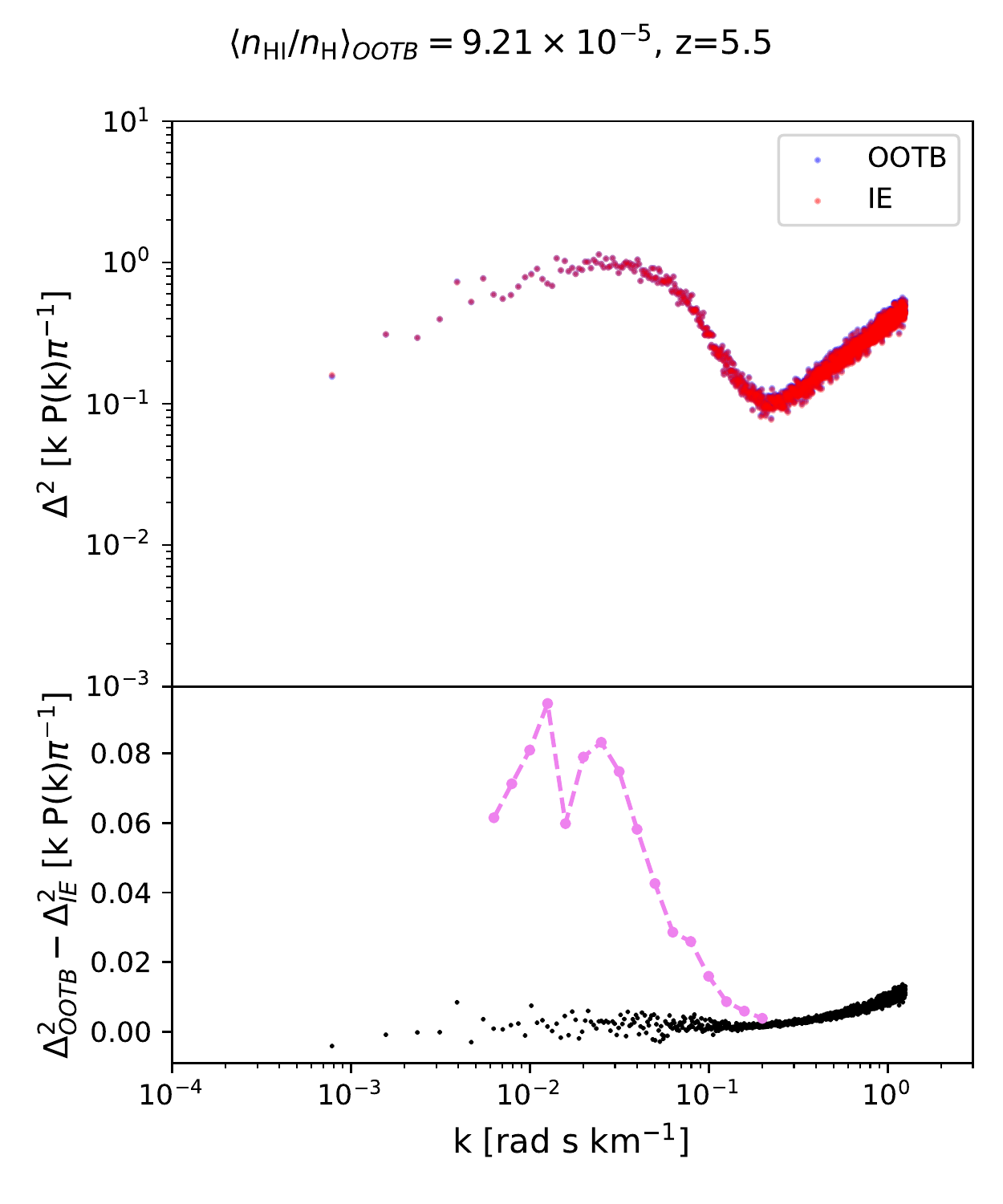}
    \caption{An extension of Figure \ref{fig:PS_almost_all} looking specifically at the snapshot at $z=5.5$. We still see the reduced power in the ionization equilibrium case, but it seems to normalize the power in small-scale structures between the two cases as we approach $z=5$.}
    \label{fig:PS_030}
\end{figure}

The previous sections demonstrated the physical differences that arise between the OOTB and IE cases. We now use these results to understand the observational consequences for the IGM. 

We begin by comparing in Figure \ref{fig:laf_spectra} normalized Lyman-$\alpha$ forest (LAF) spectra take from the two cases at $z=7$ and $z=5$. Slight differences are noticeable with more transmission in the IE case. However, it is not a uniform increase of amplitude throughout the LAF: the increase is particularly strong in regions where transmission is already high in the OOTB case. This is broadly consistent with the tendency for the IE case to create more ionized gas, particularly in voids (cf.\ Table \ref{tab:fHI_ratios}), but as the effect is inhomogeneous, it is represented only qualitatively in Figure~\ref{fig:laf_spectra}.

In order to capture the fluctuations in the transmission throughout the LAF, we will use the LAF flux power spectrum. We follow \cite{lukic15_LAF-in-sims} by defining the power spectrum along a line-of-sight for a normalized flux $\delta = \frac{f}{\langle f \rangle} - 1$ of a quasar spectrum:

\begin{equation}
    \hat{\delta}_{\mathrm{1D}}(k_{\parallel}) = L^{-1} \int \delta(x_{\parallel}) \exp{(ik_{\parallel}x_{\parallel})} \,dx_{\parallel}
\end{equation}
\begin{equation}
    \Delta^2(k_{\parallel}) \equiv \frac{k_{\parallel}}{\pi} P_{\mathrm{1D}}(k_{\parallel})
    =  \frac{k_{\parallel}}{\pi} L \braket{\hat{\delta}_{\mathrm{1D}} \hat{\delta}^*_{\mathrm{1D}}} 
\end{equation}
\noindent
Here, $k$ is the wavenumber describing the spatial quantity, $k_{\parallel}$ is computed along the line of sight, $L$ is the length of the simulated sightline in velocity space, and $\braket{\hat{\delta}_{\mathrm{1D}} \hat{\delta}^*_{\mathrm{1D}}}$ is the product of the Fourier transform and its complex conjugate shown in that order. 

To perform this, we first cast a synthetic LAF sightline through our simulation volume as described in Section~\ref{ssec:analysis}. We then compute the normalized residual flux $\delta_i$ at each pixel $i$ from the local transmission $T_i$ and the sightline-averaged mean transmission $\langle T \rangle$, which is computed over all $N$ pixels along the sightline: $\delta_i = \frac{T_i}{\langle T \rangle} - 1$ where $\langle T \rangle = \frac{1}{N}\sum_i^N{ \exp[\ln{(-A|T_i|)}}]$ for the $i$th transmission data point of a total $N$ data points and a correction factor $A$. Since we are strictly working within the simulated space and comparing the OOTB and IE cases under the assumption that the UVBs are the same, we set $A=1$. With similar neutral fraction between the cases at $z<6$, we do not expect a large difference of mean flux between the cases post-reionization. Nevertheless, much of the difference between the OOTB and IE cases may reflect \emph{slight} differences in the reionization histories because the IE case accelerates ionizations. The residual flux is binned into 8000 ``chunks." Within each 125 km s$^{-1}$ chunk, we compute the Fourier transform using the FFT library provided by the \textsc{numpy} module \citep[][]{harris2020array}. The final power spectrum is obtained by averaging over the power spectra obtained for the individual chunks.

We show the resulting power spectra at four redshifts in the IE and OOTB cases in Figure \ref{fig:PS_almost_all}. At $z=8$ there are negligible differences between the power spectra, which is somewhat surprising considering the differences of the neutral fractions between the two cases (Figure~\ref{fig:mean_fHI}). We suspect that a dominant, if not primary, cause are over-saturated damping wings from the absorption features in the LAF extending over ionized regions when represented in velocity space. This will explain why this is only applicable at $z=8$. The power spectra are similar at all redshifts, indicating that the signature of non-equilibrium ionizations in the LAF is overall weak. At large scales $k < 0.1$ rad s km$^{-1}$, there is systematically more power in the IE case during the interval $z=6$--7, reflecting the tendency for IE to accelerate reionization and temporarily enhance fluctuations in the IGM opacity. These differences vanish by $z=5$, as expected given that the volume-averaged neutral fractions converge (Figure ~\ref{fig:mean_fHI}).

A more detailed comparison, however, uncovers additional differences. Figure \ref{fig:PS_almost_all} and the bottom of Figure~\ref{fig:PS_030} show the differences between the power spectra of the OOTB and IE cases. We see from them that the LAF under ionization equilibrium has less power within the smaller-scale structures. This once again relates to the ionization fraction: as shown in Figure~\ref{fig:HI_6-5}, the OOTB case yields larger scatter in the relationship between neutral fraction versus density, particularly in the dense regions that are the last to be reionized.The tendency for the IGM under the IE assumption to have less neutral hydrogen in these small regions reflects as a reduction in the power spectrum at small scales.

The most dramatic evolution in the difference between the IE and OOTB cases occurs in the immediate aftermath of reionization, during the range $6 < z < 5$, as ionization fronts finally propagate into filaments. The disappearance of small neutral structures and consequent suppression of small-scale power occurs quickly: Figure~\ref{fig:PS_030} shows that the differences between the IE and OOTB cases have largely disappeared by $z=5.5$, in parallel with the convergence of their respective ionization fractions as seen in Figure~\ref{fig:mean_fHI}.

\section{Discussion} \label{sec:discuss}

The goal of this study was to evaluate to what extent departures from ionization equilibrium have observable consequences. We have shown that the discrepancy between the power spectra predicted in the IE and OOTB cases grows from $z=7$ to $6$ where the volume-weighted neutral fraction is $\sim 10^{-1}$ to $10^{-4}$. The discrepancy peaks around $z=6$, exceeding 4\% at $k=1$ rad s km$^{-1}$. Afterwards, it shrinks, growing negligible by $z=5$. 

In order to assess whether the discrepancy is observationally important, we compare in Figures~\ref{fig:PS_almost_all}--\ref{fig:PS_030} the magnitude of the error introduced by IE versus observational uncertainties from the state-of-the-art HIRES and UVES power spectrum measurements reported by \citep{boera19_PS-constraints}. In the bottom portion of each panel, we use a purple curve to show the reported fractional uncertainty at $z=5$. Based on this comparison, we do not expect artefacts of the assumption of IE to be detectable on velocity scales larger than 0.1 rad sec km$^{-1}$ or at neutral fractions smaller than $10^{-4}$. At scales of 0.1--0.2 rad s km$^{-1}$, the predicted offset at $z=6$ is comparable to observational uncertainties that are reported at $=5$.  Were an observational analysis of the Lyman-$\alpha$ forest flux power spectrum possible at $z>5$, however, the associated uncertainties would certainly be larger than at $z=5$. During the immediate aftermath of reionization ($z = 5.5$ in our model), we find that the effects of non-equilibrium ionizations grow nearly observable. They would be more significant at even higher resolution; \cite{boera19_PS-constraints} were only able to observe up to scales larger than $\log{k} = -0.7$, which is larger than the scale where IE introduces offsets that exceed $\sim1\%$.

Our qualitative results regarding the tendency for IE to accelerate reionization are robust to its timing, but the predicted observability of non-equilibrium effects as a function of redshift is affected in the sense that, if our model completes reionization too quickly, as suggested by a recent analyses~\citep[][]{keating20_Lya_troughs, qin2021_reion_gal-inference_LAF}, then departures from IE will be observationally relevant at lower redshifts than predicted. More concretely: our simulation predicts that reionization completes at $z=6.1$~\citep{finlator20_CIV-z>5}. If, in reality, it completes at $z<5.6$~\citep{qin2021_reion_gal-inference_LAF}, then all predictions should be referenced to a redshift that is lower by at least $\Delta z>0.5$, prolonging the interval during which non-equilibrium effects remain significant. Additionally, this work (and, in particular, Figure~\ref{fig:mean_fHI}) understates the consequences of assuming IE because it is imposed in post-processing; we would expect the tendency for IE to accelerate reionization to be stronger if we replaced our non-equilibrium ionization solver with IE and re-ran the simulation.

One possible concern with our simulations is that, as the radiation transport solver does not completely resolve ionization fronts spatially, it may artificially boost cooling and hence recombinations within partially-ionized gas. Previous work suggests that a grid cell size of $10$ $\mathrm{pkpc}$ is required to sufficiently resolve the ionization front \citep[][]{Daloisio2019}. In order to evaluate how inaccurate temperatures could impact our results, we compared the OOTB case against an isothermal test case using a simulation that subtends 6 $h^{-1}$ cMpc but is identical in all other respects to our baseline 15 $h^{-1}$ cMpc volume. The isothermal case artificially sets all gas with H number density $\nh > 0.01\:\mathrm{cm^{-3}}$ to have a single constant temperature $T=10^4\:\mathrm{K}$ in post-processing, and it leaves the UVB unchanged. We then computed power spectra from sightlines cast in the OOTB and isothermal cases and compared.

This comparison revealed that there is no significant impact on the predicted power spectrum due to temperature fluctuations from our resolution until $z \geq 6$. At $z=6$, we found a very significant decrease in power in the isothermal case with the differences at $\sim 0.7\:\mathrm{rad\:s\:km^{-1}}$ ($\sim 90\:\mathrm{ckpc}$), which is roughly the spatial width of an RT grid cell. At smaller scales, the power of the isothermal case increased until it surpassed the power of the OOTB case. Qualitatively, forcing denser regions in the IGM to be cool enhances their recombination rates, boosts their neutral fractions, suppresses thermal broadening~\citep{Peeples2010}, and, for all these reasons, increases the amount of power in small-scale fluctuations.

Although temperature fluctuations do influence the power spectrum at scales corresponding to our resolution, their impact is not significantly larger than the differences between the OOTB and IE cases. At similar scales, the impact of temperature fluctuations on the power spectrum is only about a factor of 2 greater than the assumption of ionization equilibrium. Within the context of our experiment, the OOTB and IE cases have identical temperature fluctuations, so their predictions should be impacted roughly equally.

There are several other physical effects which modulate the power spectrum at smaller scales whose significance we did not discuss.

Collisional ionization of neutrals by free electrons contributes to reionization at some level, reducing the power in the LAF power spectrum beyond what is expected in models that assume only photoionization \citep[][]{rahmati16_highly-ionied-metals-dist}. This effect is already accounted for in both the OOTB and IE predictions. We assessed its significance using a trial simulation that subtends 6 h$^{-1}$ cMpc but is in all other respects identical to our baseline, 15 h$^{-1}$ cMpc simulation. We considered three cases: OOTB, IE, and IE but without collisional ionizations. The largest effect was seen at scales of $0.1\:\mathrm{rad\:s\:km^{-1}}$, where fractional differences grew to 1.5\%. This is small compared to the already-small difference between IE and OOTB. We conclude that collisional ionizations do not impact the small-scale LAF power spectrum significantly. 

The length scale where non-equilibrium effects are strongest is similar to the length scale where galactic feedback and contaminating metal absorbers are expected to modulate the LAF power spectrum. Our simulations do model feedback, but we do not evaluate the impact of feedback by removing it because doing so would drastically change the predicted reionization history. Metal-line cooling has much weaker effects on the power spectra compared to galactic outflows; \cite{viel13_gal-feedback-laf} show this as well as the reduction in power due to AGN and SNe feedback being comparable, where $ k > 0.1\:\mathrm{s/km}$ at $z=5$ is comparable to the increase of SNe contaminating the power at $k > 15\:\mathrm{h/Mpc}$ at $z=4$. \cite{chabanier20_AGNfeedback_PS} likewise report a weak effect from AGN feedback. Additionally, both studies do show a decrease in the importance of galactic outflows at higher redshifts, where the differences in the power spectra are less than that of ionization equilibrium; however, both studies only went up to $z=4$ and $z=4.25$, respectively. It would be useful to extend their study to higher redshifts in order to evaluate the impacts of feedback and metal absorber contamination on the reionization-epoch LAF.

Another effect that has significant impact on the power spectrum is temperature fluctuations, specifically pressure smoothing and thermal broadening. Significant deviations from both effects occur at $5.4 \leq z \leq 4.9$ \citep[][]{nasir2016_thermal-history-z5, wu2019_temp-flucs-EOR, Peeples2010}, growing up to a 20\% fractional difference at similar scales to ours. The simulation accounts for pressure smoothing and thermal broadening, and they are taken into account automatically by the sightline generation. However, we did not study either effect, as our focus was on ionization equilibrium. Comparing our results to theirs, we see that non-equilibrium effects are overall weaker than temperature fluctuations. However, for a potential future project, it will be important to further look into how temperature fluctuations deeper into cosmic time, such as where we see our biggest impact from non-equilibrium effects, at and shortly after overlap.

Our IE model is not self-consistent because ionization equilibrium is imposed during post-processing. Our results show the UVB outputted directly from our simulation runs, which utilize non-equilibrium modelling. If we re-ran our calculations under the assumption of ionization equilibrium, the reduced neutral fractions throughout, especially in void regions, would increase the mean free path of ionizing photons and potentially further ionize the universe. This will accelerate reionization and potentially suppress galaxy formation. Although this is potentially a significant effect, our tests were designed to see how much of an impact assuming ionization equilibrium has if one was to assume it during observations, regardless of present gas properties. Future work can be done to see impacts on a universe depending on the cases, such as with galactic evolution and IGM properties.

\section{Conclusions} \label{sec:conclusion}

We explore the differences in physical properties and LAF power spectrum of a simulated IGM as predicted by a self-consistent reionization model versus a case in which the ionization equilibrium is imposed in post-processing. Our main conclusions are as follows:
\begin{itemize}
    \item Regions with higher neutral fractions have reduced UVB amplitudes and  photoionization rates. This is mainly seen as a large opacity around the $\mathrm{HI}$ ionization threshold.
    \item IE artificially accelerates reionization, with the result that the IGM under IE is considerably more ionized within the range $8 \leq z \leq 5$. Additionally, IE introduces a maximum neutral fraction wherever $\log{(\Delta_{\mathrm{H}})} < 1$ and this maximum decreases as $\Delta_{\mathrm{H}}$ decreases.
    \item Ionization equilibrium artificially suppresses scatter in the ionization and recombination timescales, photoionization rates, and neutral fractions.
    \item Owing to the tendency for IE to accelerate reionization by exaggerating the ionization rate within ionization fronts, IE suppresses power in the LAF in small-scale structures within the range $7 \leq z \leq 5.5$. Although this may not be currently detectable with present technology, it may be in the future.
\end{itemize}

However, improvements can be done in order to get more robust results, or potentially more different results. Our simulation box size is relatively small, but capable to show the difference in the power spectra due to non-equilibrium effects. One can increase the box size and increase spatial resolution in order to study the power spectra in more detail, or to even study other potential observables that we cannot due to our resolution limits. As such, increasing the dynamic range of the simulations could increase the fidelity of observables in such simulation, leading to possibly higher resolved data.

It is possible to extend this research's results into observations. Our simulated spectra follow an instrument's response FWHM of 6 km/s. We expect any future telescope on the ground that is 30m-class or larger with appropriate instrumentation to be able to retrieve data in a similar resolution as we simulated, if not better \citep[for example, MODHIS;][]{mawet2019_MODHIS}. As such, observations may grow sensitive to non-equilibrium effects within the coming years. 

\begin{acknowledgments}
We want to thank Caitlin Doughty for reviewing our draft and improving the writing. Samir Ku\v{s}mi\'{c} is supported by the National Science Foundation (NSF) under Award Number 2006550. The Technicolor Dawn simulations were enabled by the Extreme Science and Engineering Discovery Environment (XSEDE), which is supported by NSF grant number ACI-1548562. The Cosmic Dawn Center is funded by the Danish National Research Foundation. Laura Keating was supported by the European Union’s Horizon 2020 research and innovation programme under the Marie Skłodowska-Curie grant agreement No. 885990. This article is under the Creative Commons Attribution 4.0 International license.
\end{acknowledgments}

\section*{Data Availability}
The data used for this research are available from the authors upon request.

\bibliography{export-bibtex}{}
\bibliographystyle{aasjournal}



\end{document}